\documentclass[twocolumn]{extarticle} %'ext' is so we can get larger base font sizes
\usepackage[utf8]{inputenc}

\title{Relaxed phase-matching constraints in zero-index waveguides}

\newsavebox{\foobox}
\newcommand{\slantbox}[2][0]{\mbox{%
        \sbox{\foobox}{#2}%
        \hskip\wd\foobox
        \pdfsave
        \pdfsetmatrix{1 0 #1 1}%
        \llap{\usebox{\foobox}}%
        \pdfrestore
}}
\newcommand\unslant[2][-.25]{\slantbox[#1]{$#2$}}
\newcommand{\micron}{$\unslant\mu$\rm{m}}

\usepackage[margin=0.7in]{geometry}
\usepackage{titling} %For Supp Materials

\usepackage{amsmath}
\usepackage{amssymb}
\usepackage{graphicx} % use png / jpg images
\usepackage{caption} % The following format will be used to emulate the captions
\usepackage{float}    % use [htb] float location, or H (stricter than h!)
\usepackage[colorlinks=true, allcolors=blue]{hyperref}
\usepackage{cite}
\usepackage{authblk} % for author affiliations
\usepackage{textgreek} % insert Greek letters
\usepackage{xcolor} % change text color

\usepackage[resetlabels]{multibib}
\newcites{SM}{References for Supplementary Information}

\DeclareCaptionLabelFormat{adja-page}{\hrulefill\\#1 #2 \emph{(previous page)}}
\usepackage{subcaption}

\newcommand{\SI}[1]{Sec.~\ref{#1}:~\nameref{#1}}

\DeclareMathOperator{\sinc}{sinc}

\author[1,$\dagger$]{Justin~R.~Gagnon}
\author[1,*,$\dagger$]{Orad~Reshef}
\author[2]{Daniel~H.~G.~Espinosa}
\author[1]{M.~Zahirul~Alam}
\author[3]{Daryl~I.~Vulis}
\author[3]{Erik~N.~Knall}
\author[1]{Jeremy~Upham}
\author[3,4]{Yang~Li}
\author[1,2]{Ksenia~Dolgaleva}
\author[3]{Eric~Mazur}
\author[1,2,5]{Robert~W.~Boyd}
\affil[1]{Department of Physics, University of Ottawa, 25 Templeton Street, Ottawa, ON K1N 6N5, Canada}
\affil[2]{School of Electrical Engineering and Computer Science, University of Ottawa, 25 Templeton Street, Ottawa, ON \ K1N 6N5, Canada}
\affil[3]{John A. Paulson School of Engineering and Applied Sciences, Harvard University, 9 Oxford Street, Cambridge, Massachusetts 02138, USA}
\affil[4]{State Key Laboratory of Precision Measurement Technology and Instrument, Department of Precision Instrument, Tsinghua University, 100084 Beijing, China}
\affil[5]{Institute of Optics and Department of Physics and Astronomy, University of Rochester, 500 Wilson Blvd, Rochester, New York 14627, USA}
\affil[$\dagger$]{J. Gagnon and O. Reshef contributed equally to this work.}
\affil[*]{Corresponding author: orad@reshef.ca}

\date{\vspace{-2em}}

\begin{document}

\twocolumn[
  \begin{@twocolumnfalse}

\maketitle

\begin{abstract}
\noindent The nonlinear optical response of materials is the foundation upon which applications such as frequency conversion, all-optical signal processing, molecular spectroscopy, and nonlinear microscopy are built~\cite{garmire2013,willner2014,mukamelnonlinearspectroscopy,schermelleh2010}. 
However, the utility of all such parametric nonlinear optical processes is hampered by phase-matching requirements~\cite{boydnonlinearoptics}.
Quasi-phase-matching~\cite{armstrong1962,yamada1993}, birefringent phase matching~\cite{midwinter1965}, and higher-order-mode phase matching~\cite{evans2015,levy11} have all been developed to address this constraint, but the methods demonstrated to date suffer from the inconvenience of only being phase-matched for a single, specific arrangement of beams, typically co-propagating, resulting in cumbersome experimental configurations and large footprints for integrated devices~\cite{suchowski2013}.
Here, we experimentally demonstrate that these phase-matching requirements  may be satisfied in a parametric nonlinear optical process for multiple, if not all, configurations of input and output beams when using low-index media.
Our measurement constitutes the first experimental observation of direction-independent phase matching for a medium sufficiently long for phase matching concerns to be relevant.
We demonstrate four-wave mixing from spectrally distinct co- and counter-propagating pump and probe beams, the backward-generation of a nonlinear signal, and excitation by an out-of-plane probe beam.
These results explicitly show that the unique properties of low-index media relax traditional phase-matching constraints, which can be exploited to facilitate nonlinear interactions and miniaturize nonlinear devices, thus adding to the established exceptional properties of low-index materials~\cite{reshef2019}.

% Start word-counting after this
%TC:endignore

\end{abstract}

    \noindent\hfil\hspace{1cm}\rule{0.8\textwidth}{.4pt}\hfil
    \vspace{1em}
  \end{@twocolumnfalse}
]

When light is generated by a parametric nonlinear interaction (\emph{e.g.,} harmonic generation~\cite{franken1961}), the propagation direction of the generated output light is dictated by the properties of the input beams~\cite{boydnonlinearoptics, suchowski2013}. This dependence is due to conservation of momentum, also known as phase-matching~\cite{boydnonlinearoptics,agrawalnonlinearfiberoptics}. The amount by which the phase-matching condition is not satisfied is quantified by the phase mismatch, $\Delta k$, the difference in the momentum of the constituent beams. Approaches such as quasi-phase-matching~\cite{armstrong1962,yamada1993}, birefringent phase matching~\cite{midwinter1965}, and higher-order-mode phase matching~\cite{evans2015,levy11} have been demonstrated as means to achieve phase matching. However, these methods suffer from the inconvenience of only being phase-matched for one specific configuration of the participating beams, which is typically collinear and along the direction of propagation~\cite{suchowski2013}, and only for a narrow range of wavelengths~\cite{lan2015}. These constraints pose severe limitations on potential applications in nonlinear optics, where flexibility and compactness are highly desired.

There has been significant interest in using metamaterials to lift such constraints and explore the resulting novel behavior~\cite{lan2015,suchowski2013,luo2020,volochbloch2010,planat2020,kinsey2019,wang2017}. Metamaterials provide ultimate flexibility in the engineering of optical materials, enabling many unusual and interesting properties, including negative indices of refraction~\cite{shelby2001,valentine2008,lezec2007}. Materials with a negative refractive index have been used to demonstrate the second-harmonic generation of a nonlinear signal wave propagating against the pump wave, known as backward phase matching~\cite{lan2015,liangliang2018}. This unique behavior may be further explored when considering zero-index media~\cite{liberal2017,vulis2018}.

\begin{figure}[ht!]
    \includegraphics[width=\linewidth]{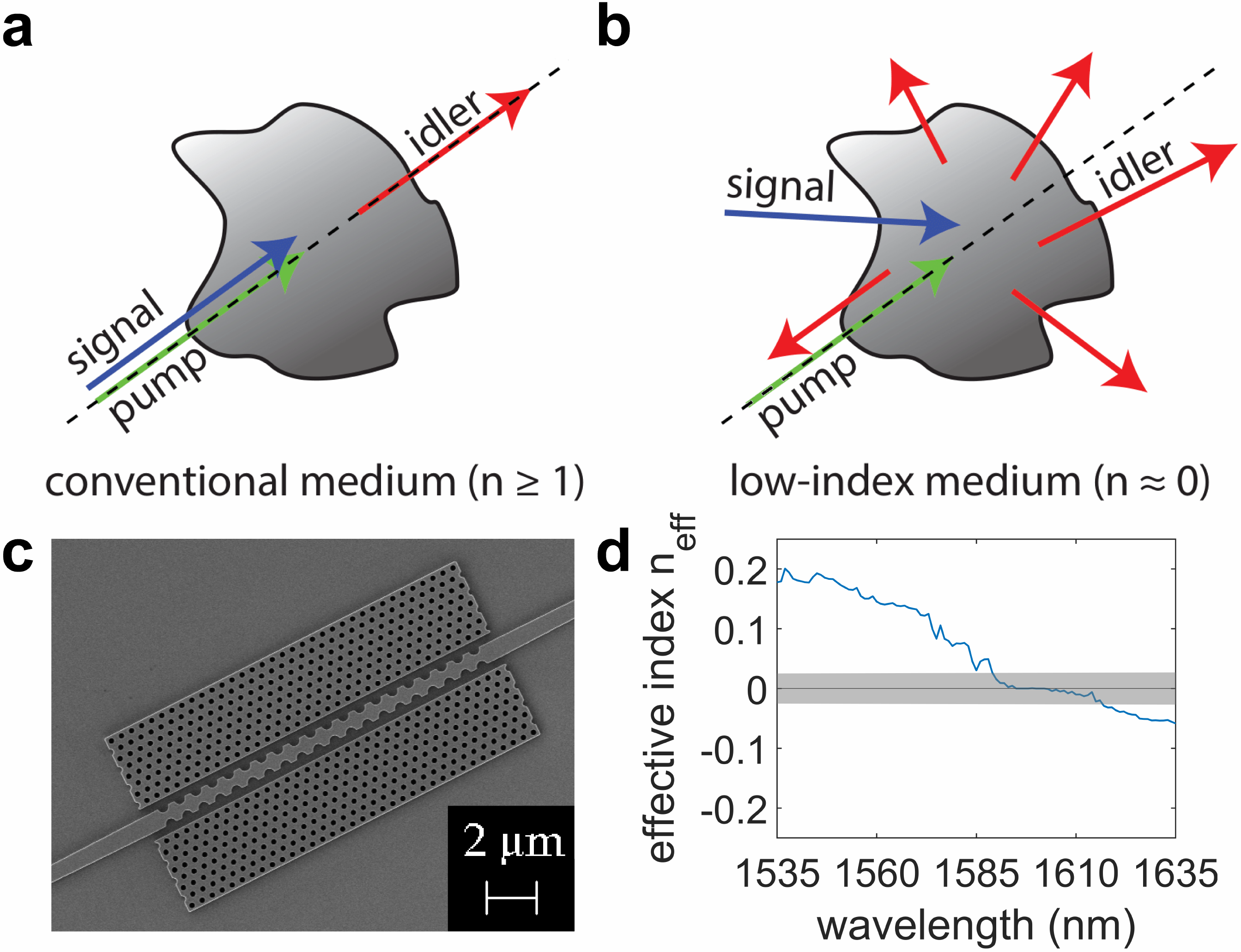}
    \caption{{\bf Phase matching in a low-index medium.} {\bf a)~}In a conventional medium ($n>1$), the input and output beams must be carefully aligned, typically co-propagating, to satisfy the phase-matching condition. In a four-wave mixing interaction, this corresponds to aligning the signal beam with the pump beam to generate a collinear idler beam. {\bf b)~}In a low-index medium  ($n\approx 0$), the constituent beams are free to adopt any orientation and still maintain phase-matching. {\bf c)~}Scanning electron microscope image of a Dirac-cone zero-index waveguide surrounded by photonic band gap materials (triangular lattice of holes). {\bf d)~}Refractive index profile of one of the waveguides used in the experiment, crossing zero at $\lambda = 1600$~nm. The shaded region indicates a refractive index below the measurement threshold of $n< 0.02$.}
    \label{fig:chip+N_profile}
\end{figure}

As the magnitude of the momentum wave-vector $k$ is proportional to the refractive index $n$ ($k = 2\pi n / \lambda$, where $\lambda$ is the free-space wavelength), it vanishes for light propagating in a zero-index medium. Consequently, light in a zero-index mode does not contribute any momentum to phase-matching considerations, and its propagation direction becomes inconsequential to the phase mismatch (Figs.~\ref{fig:chip+N_profile}a~--~b). By virtue of this unique quality, many otherwise forbidden phenomena, such as the simultaneous generation of both forward and backward-propagating light, become possible~\cite{suchowski2013}.

In our experiment, we explore these phenomena using Dirac-cone metamaterials that achieve an effective refractive index of zero via the simultaneous zero-crossing of the permittivity and permeability while maintaining a finite impedance~\cite{vulis2018,reshef2017}. These metamaterials consist of a pair of silicon-based, corrugated ridge waveguides whose dispersion profiles have zero-crossings at 1600~nm or 1620~nm. Figures \ref{fig:chip+N_profile}c,d show an image of a fabricated waveguide and its measured refractive index profile. By sampling five distinct configurations of pump, signal, and idler waves, our experimental results support the existence of \emph{direction-independent} phase matching (See \SI{sec:unrestricted}).

In the four-wave mixing (FWM) interaction under investigation, a powerful pump beam interacts with a signal (probe) beam, converting two pump photons of frequency $\omega_p$ into one signal photon of frequency $\omega_s$, and one idler photon of frequency $\omega_i = 2\omega_p - \omega_s$~\cite{foster2006}. 
As is usually studied, all the beams of a FWM process are co-propagating, and the phase mismatch is given by $\Delta k_\mathrm{fw} = 2k_p - k_s - k_i$, where suffixes \emph{p}, \emph{s}, and \emph{i} represent the pump, signal, and idler, respectively. In a standard silicon ridge waveguide, this phase-matching condition may be satisfied ($\Delta k \approx 0$). However, the phase mismatch would then prevent efficient FWM if the idler wave traveled in the backward direction (\emph{i.e.,} counter-propagating with respect to the pump beam), because then $\Delta k_\mathrm{bw} = 2k_p - k_s + k_i = \Delta k_\mathrm{fw} + 2k_i \approx 2k_i$. Similarly, phase mismatch would prevent efficient FWM if the signal wave was counter-propagating against the pump wave.

To explore the impact of a low-index response on phase matching, we consider the special case of co-propagating input beams when the idler wave is generated at the zero-index wavelength. In this case, any generated nonlinear forward and backward-propagating signal would be expected to increase with equal efficiency due to the vanishing momentum contribution of $k_i$. Indeed, simulations predict that the backward-propagating idler wave is strongest when the idler is located at the zero-index wavelength (See \SI{sec:simulations}).

\section*{Results}

\begin{figure}[ht!]
    \includegraphics[width=\linewidth]{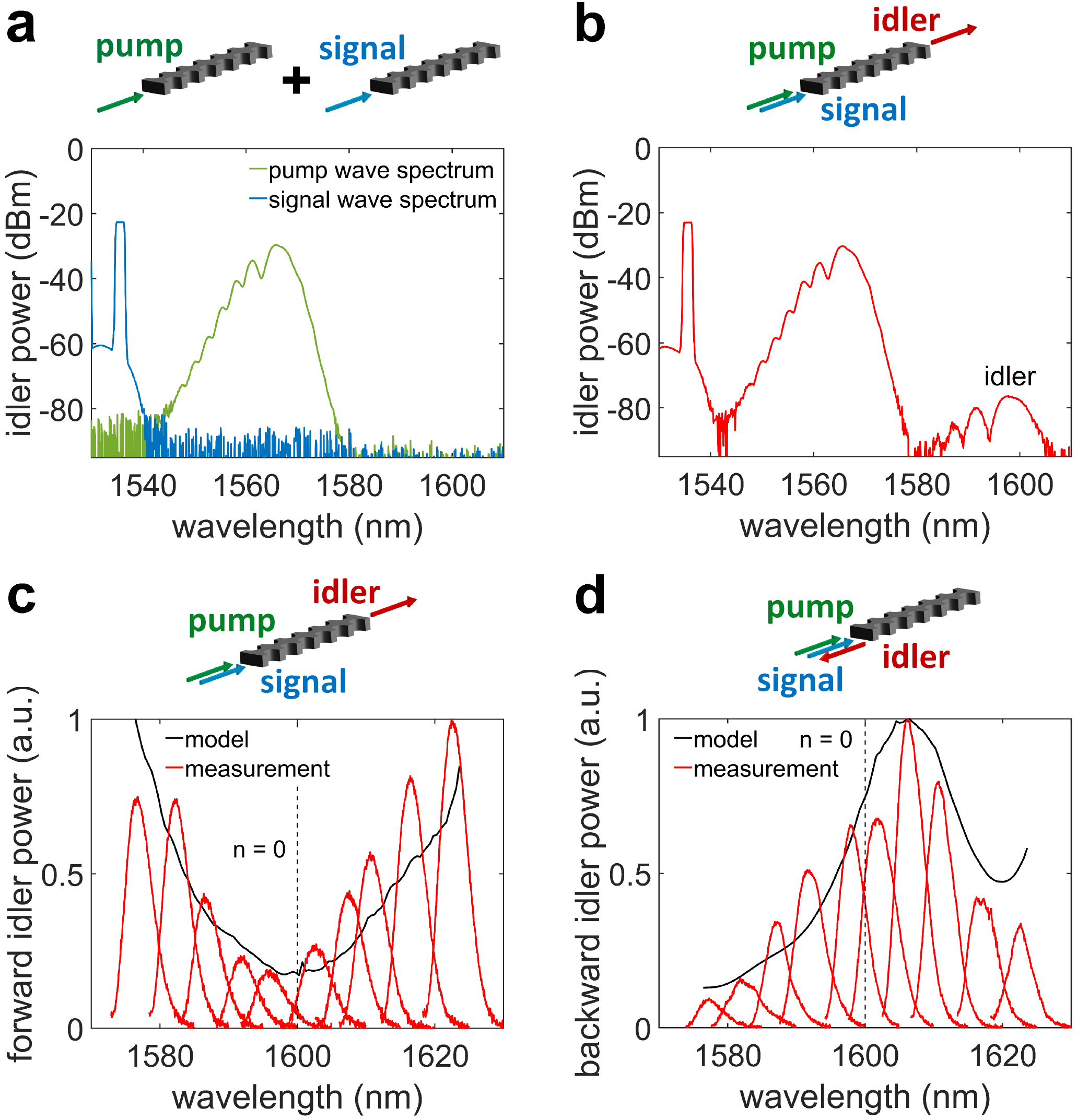}
    \caption{{\bf Collinear phase-matching measurements.} {\bf a)~}Spectra of the pump and signal waves when measured after propagating independently through a 15-\micron{}-long low-index waveguide. {\bf b)~}When these same pump and signal beams are simultaneously applied to the waveguide, an idler wave is generated in the forward direction at $\omega_i = 2\omega_p - \omega_s$ (1600~nm). The spectrum of the idler wave closely follows that of the pump wave because of the narrowness of the signal-beam spectrum. Generated idler wave spectra in the {\bf c)~}forward and {\bf d)~}backward directions in a FWM process with co-propagating pump and signal beams. The red curves show the spectra of the idler beams for ten different values of the pump and signal wavelengths (See Methods). For each wavelength pair, the spectral gap between the pump and signal frequencies is held constant. The black curves show the peak power of the pulses predicted by phase-matching constraints (See \SI{sec:loss_derivation}), while the vertical dotted black lines in {\bf (c)} and {\bf (d)} indicate the $n=0$ wavelength.}
    \label{fig:example_spectra_backward_forward}
\end{figure}

As a first step towards demonstrating directionally unrestricted phase matching, we show the simultaneous generation of forward and backward-propagating idler waves when considering the pump and signal beams co-propagating in a waveguide (Fig.~\ref{fig:example_spectra_backward_forward}). Through the careful simultaneous adjustment of the pump and signal beams, this measurement produces idler waves for wavelengths of $\lambda_i$ ranging from 1570 to 1630~nm, crossing through the zero-index wavelength at $\lambda = 1600$~nm (Fig.~\ref{fig:example_spectra_backward_forward}c~--~d). The backward-propagating light peaks at $\lambda_i$ = 1606~nm, while the forward-propagating light has a dip centered at 1596~nm. We also plot our theoretical predictions alongside our experimental results (black curves in Figs.~\ref{fig:example_spectra_backward_forward}c~--~d). The forward and backward-generated spectra show almost perfect agreement with the theory in terms of both peak wavelength and rate of drop off. The forward-generated idler wave dips in power shortly before the zero-index wavelength at 1596~nm. This dip is caused by dispersive propagation loss and permeability values (See \SI{sec:loss_derivation}). Such effects are less prominent in the backward-generated light, where phase matching is shown to be the dominant factor~\cite{reshef2017}. Beyond the strong theoretical agreement, the fact that the most powerful backward-generated idler wave is not located at the same wavelength as the least powerful forward-generated idler wave constitutes additional proof that the backward-propagating idler wave is independently generated, and does not merely consist of back-scattering of the forward-propagating light due to possible reflections at the zero-index wavelength.

\begin{figure}[ht!]
    \includegraphics[width=\linewidth]{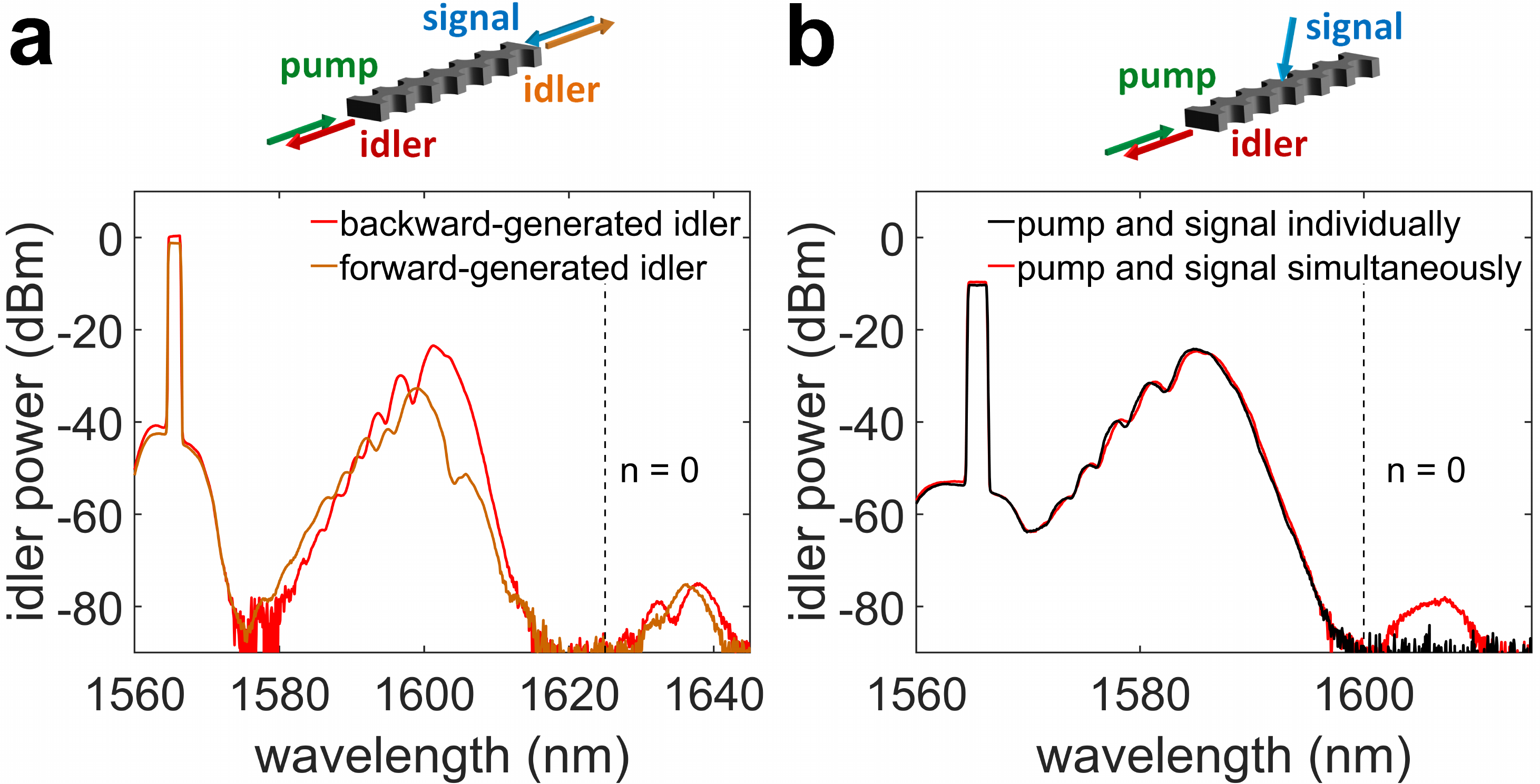}
    \caption{{\bf Counter-propagating and out-of-plane phase-matching measurements.} {\bf a)~}Spectra showing FWM for counter-propagating pump and signal beams with an idler wave generated in the forward (co-propagating with the pump wave, orange) and backward (co-propagating with signal wave, red) directions. The signal and pump beams are at 1565~nm and 1600~nm, respectively, while the idler wave appears at 1635~nm. {\bf b)~}Generated idler wave spectrum resulting from a signal beam coupling from out of the plane of the waveguide. An idler wave is generated in the backward direction only when the pump and signal beams are simultaneously applied (red curve compared to the blue curve). The vertical dotted black lines in both figures indicate the $n=0$ wavelength.}
    \label{fig:co-propagating_oop}
\end{figure}

We next consider the phase-matching condition for other phase-matching configurations not possible in conventional waveguides. For counter-propagating pump and signal beams, simulations predict that the brightest forward-propagating idler wave will occur when the signal wave is at the zero-index wavelength (here at $\lambda$~=~1620~nm), while for the backward-propagating idler wave it is predicted when the pump wave is at the zero-index wavelength. We perform measurements with a pump beam at 1600~nm and a signal beam at 1565~nm, where both requirements are best satisfied given experimental limitations. The resulting spectra are shown in Fig.~\ref{fig:co-propagating_oop}a. The simultaneous generation of forward and backward-propagating idler waves is again clearly visible, here at $\lambda_i$~=~1630~nm.

We further establish phase matching without directional restriction by coupling the pump beam into the waveguide as before and shining the signal beam onto the waveguide from out of the plane of the device. A backward-propagating idler wave is observed at $\lambda_i$~=~1605~nm as shown in Fig.~\ref{fig:co-propagating_oop}b. In addition to confirming our theoretical predictions, observing the FWM process from a signal beam coupling from outside the plane of the device layer provides further proof that low-index waveguides significantly ease restrictions on parametric nonlinear effects by relaxing the phase-matching condition.

\section*{Discussion}

The simultaneous generation of forward and backward-propagating idler light has been previously observed in a fishnet metamaterial with a total thickness of 800~nm~\cite{suchowski2013}. However, the thickness of that metamaterial was smaller than the free-space optical wavelength ($\lambda = 1510$~nm) and phase mismatch is not a concern over such small propagation lengths~\cite{kauranen2012}. Our demonstration uses similar wavelengths but a 14.8~\micron{} long waveguide, corresponding to almost 10 free-space optical wavelengths and consistent with a lower bound estimate of the coherence length at 7.8~\micron~(See \SI{sec:lower_bound}). Therefore, low-index waveguides address the phase mismatch challenge rather than side-stepping it, as would be the case in a thin metasurface configuration. In addition, while this earlier demonstration used intra-pulse FWM, our demonstration uses multiple spectrally-distinct beams, enabling the clean isolation of the generated nonlinear pulses from the inputs, resulting in an  unambiguous demonstration. These factors support the conclusion that the process is strongly phase-matched. While our current zero-index platform exhibits radiative losses, some methods have been proposed to reduce loss in similar zero-index platforms~\cite{camaydmunoz2016,dong2021,tang2021,momchil2018}.

Observation of the idler wave generated in the waveguide (Fig.~\ref{fig:co-propagating_oop}b) when excited from outside the device layer could be explained in two ways: 1) The signal beam, which is incident on the waveguide from outside the device and from a direction very different to that of the guided modes, can generate an idler wave because the phase-matching condition has been so relaxed by the wave vectors vanishing at at low refractive indices, as we claim; 2) The signal beam couples into the guiding mode of the waveguide from free space, and subsequently generate FWM at the low refractive indices. The present experiment cannot distinguish between these two explanations for the observed idler waves. However, it is clear that the vanishing $k$-vector, and therefore, a near-zero refractive index, is the key to enabling FWM in the waveguide when excited from outside the device layer.

In summary, we have experimentally demonstrated that a low-index medium enables phase-matching free of directional restriction for the constituent beams, which greatly relaxes conventional nonlinear optical constraints and potentially enables all input and output beams to take on any desired configuration. While low-index materials still require conventional phase matching through the careful engineering of its dispersion parameter (See \SI{sec:dispersion}), they provide great flexibility in terms of propagation direction. We believe that such structured low-index media have the potential to facilitate the realization of nonlinear optical interactions due to the relaxation of this constraint and thus serve innumerable roles in the field of nonlinear optics. 

\section*{Author contributions}
JRG carried out the nonlinear measurements. OR conceived the basic idea for this work. JRG, OR, and DHGE designed the experiment. DIV and EK carried out the linear measurements. OR and YL carried out the simulations. JRG, OR, JU, and ZA analyzed the experimental results. RWB, JU, EM, and KD supervised the research and the development of the manuscript. JRG and OR wrote the first draft of the manuscript, and all authors subsequently took part in the revision process and approved the final copy of the manuscript.

\section*{Acknowledgements}
Fabrication in this work was performed in part at the Center for Nanoscale Systems (CNS), a member of the National Nanotechnology Coordinated Infrastructure Network (NNCI), which is supported by the National Science Foundation under NSF award no. 1541959. CNS is part of Harvard University. 

The authors thank Kevin P. O'Brien for fruitful discussions. The authors gratefully acknowledge support from the Canada First Research Excellence Fund, the Canada Research Chairs Program, and the Natural Sciences and Engineering Research Council of Canada (NSERC [funding reference number RGPIN/2017-06880]). 
R.W.B. and E.M. acknowledge support from the Defense Advanced Research Projects Agency (DARPA) Defense Sciences Office (DSO) Nascent program and the US Army Research Office. O.R. acknowledges the support of the Banting Postdoctoral Fellowship of the Natural Sciences. Portions of this work were presented at the 2016 Conference on Lasers and Electro-Optics (CLEO) in San Jose, CA~\cite{reshef2016CLEO}.

\section*{Methods}
The waveguides used in the experiment were fabricated by writing a pattern into a negative-tone resist using electron-beam lithography, and subsequently transferring it to a silicon substrate using inductively-coupled plasma reactive ion etching~\cite{reshef2017}.
To facilitate coupling into the waveguides, polymer coupling pads with large cross-sectional areas were constructed on either end of the waveguide.
The waveguides consist of a row of zero-index Dirac cone metamaterial with a lattice constant of $a$ = 760~nm and a cylindrical hole of radius $r$ = 212~nm~\cite{reshef2017}.
Two zero-index waveguides are used: waveguide A with a length of 14.8~\micron{} and a zero-index wavelength of 1600~nm, and waveguide B with a length of 11.1~\micron{} with a zero-index wavelength of 1625~nm. 
Their propagation loss has been previously determined to be wavelength-dependent, with values of up to 1 dB/\micron~\cite{reshef2017}.
Waveguide A is used for the co-propagating and out-of-plane measurements (Figs.~\ref{fig:example_spectra_backward_forward} and~\ref{fig:co-propagating_oop}b).
However, in a setup featuring counter-propagating beams, there is less power overlap between the pump and signal beams as a result of propagation losses in the waveguide.
As a result, waveguide B is used for the counter-propagating measurements (Fig.~\ref{fig:co-propagating_oop}a) due to its shorter length which allows for a larger power overlap. 

In this experiment, a pulsed laser provides the pump beam, and an amplified continuous-wave laser provides the signal seed beam. The full setup can be seen in detail in \SI{sec:setup}.
The pulsed laser consists of a Ti:Sapphire and optical parametric oscillator pumped by a 532~nm continuous-wave laser. This setup is capable of generating infrared pulses with a peak power of 1300~W, a pulse width of 3~ps, and a repetition rate of 76~MHz.
The signal laser consists of a continuous-wave laser amplified by an erbium-doped fiber amplifier capable of accessing wavelengths between 1535~nm and 1565~nm with a peak power of 2~W.
In measurements with co-propagating beams involving a signal beam above 1565~nm, a weaker erbium-doped fiber amplifier capable of generating up to 100~$\unslant\mu$W was used.
The spectra exiting the waveguide are measured using an optical spectrum analyzer set to a resolution of 2~nm.

In our measurement with co-propagating pump and signal beams, we sweep the pump wavelength in increments of 5~nm from 1555~nm to 1600~nm while maintaining a constant spectral separation between the pump and signal waves ($\Delta f = c/ \lambda_p - c/ \lambda_s = 2.4$~THz). A constant spectral separation ensures dispersion will not contribute to any changes in the power of the generated idler waves. The corresponding pump and signal wavelengths used to produce the idler peaks are: 1555~nm, 1536.1~nm; 1560~nm, 1540.98~nm; 1565~nm, 1545.86~nm; 1570~nm, 1550.74~nm; 1575~nm, 1555.62~nm; 1580~nm, 1560.49~nm; 1585~nm, 1565.37~nm; 1590~nm, 1570.25~nm; 1595~nm, 1575.12~nm; 1600~nm, 1580~nm. The power of these generated peaks has been shown to vary quadratically with the power of the pump wave; this dependence confirms that the peaks are the result of a FWM interaction (See \SI{sec:power_measurement}). In our measurement with counter-propagating pump and signal beams, the deviation in the shape of the generated idler wave spectra is due to spectral changes incurred by propagation in the waveguide, as well as fluctuations in the spectrum of the pump beam.  When the signal beam is incident on the sample from outside the beam plane, the measurements are performed with the pump wave at $\lambda$~=~1585~nm and the signal wave at 1565~nm. As the waveguide will only accept light coming in at an incident angle defined by Snell’s law and the refractive index at 1565~nm is slightly positive (n $\approx$ 0.17), the signal beam is angled 9.8 degrees off normal incidence. For maximum power overlap between strong pump and signal beams, the signal beam is introduced from out-of-plane closest to the side of the waveguide where the pump beam is introduced.

%%%%%%%%%%%%%%%%%%%%%%% SUPPLEMENTAL %%%%%%%%%%%%%%%%%%%%%%%%
%%%%%%%%%%%%%%%%%%%%%%%%% MATERIALS %%%%%%%%%%%%%%%%%%%%%%%%%

\pagebreak

\onecolumn
% Restart the page, section and figure counters:
\setcounter{page}{1}
\renewcommand\thesection{S\arabic{section}} 
\setcounter{section}{0}
\renewcommand\thefigure{S\arabic{figure}}   
\setcounter{figure}{0}  

% Set the Supplementary Information title
\makeatletter
    \let\Title\@title
    \newcommand{\newtitle}{Supplementary Information: \Title}
\makeatother
\title{\newtitle}

\maketitle

This document provides Supplementary Information for ``Relaxed phase-matching constraints in zero-index waveguides.'' In Section \ref{sec:unrestricted}, we demonstrate a theoretical treatment which provides support for phase matching free of dimensional restriction. In Section \ref{sec:simulations}, we show a simulation of the generated nonlinear signal as a function of propagation length for forward and backward-propagating idler waves using nonlinear scattering theory. In Section \ref{sec:loss_derivation}, we show the model used to generate the theoretical curves in Fig.~\ref{fig:example_spectra_backward_forward}, as well as provide an explanation for the shape of the profile of the forward-generated idler peaks. In Section \ref{sec:lower_bound}, we provide our calculation of the lower bound estimate on the coherence length. In Section \ref{sec:dispersion}, we show the simulated dispersion profile of our zero-index waveguide. In Section \ref{sec:setup}, we show and provide an overview of the setup used to collect our data. In Section \ref{sec:power_measurement}, we show the data for the power measurement that was used to demonstrate that the nonlinearity is third-order FWM.

\section{Idler power predictions and theoretical support for phase matching free of directional restriction}
\label{sec:unrestricted}

It can be useful to think of phase matching in terms of the coherence length~\citeSM{mattiucci20072} given by
\begin{equation}
	L_{\mathrm{coh}} = 2 / \Delta k.
	\label{eq:Lcoh}
\end{equation}
This parameter indicates the length over which a nonlinear interaction remains coherent, \emph{i.e., }where there is constructive interference of a generated idler wave. The lower the phase mismatch, the longer the coherence length.

\begin{figure}[H]
    \includegraphics[height=1.5in]{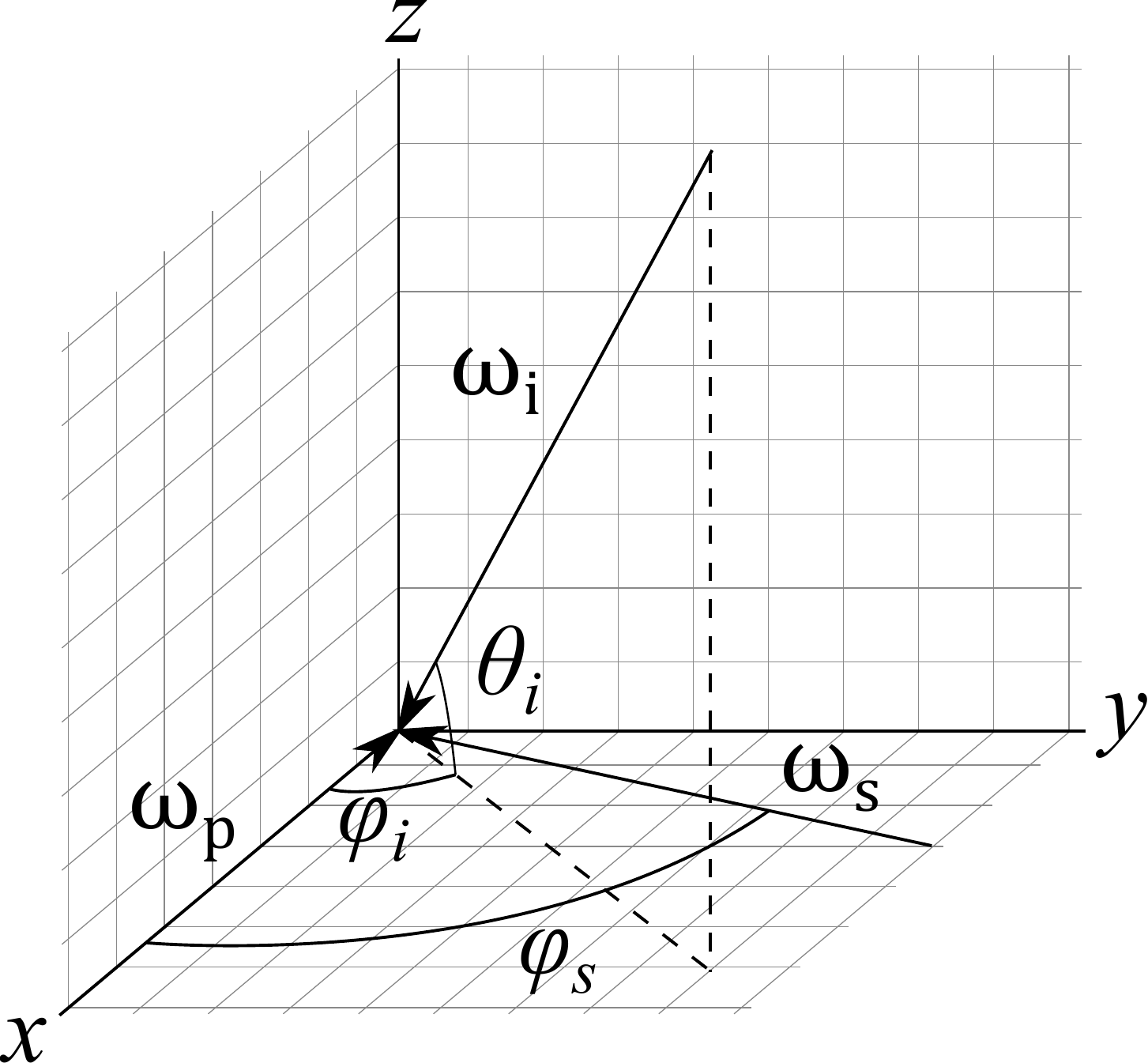}
    \centering
    \caption{{\bf Representation of the three angles $\phi_s$, $\phi_i$, and $\theta_i$ for arbitrary signal and idler waves relative to the pump wave.} Here, the pump wave is located on the x-axis, and the signal wave is located along the $xy$-plane. Note that $\theta_i$ is defined coming up from the $xy$-plane, in contrary to the regular definition with spherical coordinates.}
    \label{fig:3D_spherical}
\end{figure}

To obtain the coherence length for an arbitrary beam configuration, we can make the reasonable assumption that, as in all collinear cases, the phase relationship between the constituent beams is the principle governing factor in the generation of a powerful idler wave. Therefore, to judge the phase-matching properties of this beam configuration, we can calculate the coherence length $L_{\mathrm{coh}}$ by generalizing the phase-matching relation for all possible orientations of pump, signal, and idler waves. To do this, we split up the phase-matching relation into its $x$, $y$, and $z$ components
\begin{equation}
	|\Delta \vec{k}| = \sqrt{{\Delta k_x}^2 + {\Delta k_y}^2 + {\Delta k_z}^2},
	\label{delta_k_arbitrary}
\end{equation}
where $\Delta k_x$, $\Delta k_x$, and $\Delta k_z$ are defined by using the angles of the pump, signal, and idler waves as in Fig.~\ref{fig:3D_spherical}. Without loss of generality, we may define the pump wave as being at $\phi,\theta = 0$, where $\phi$ and $\theta$ are the azimuthal and polar angles, respectively. spherical coordinates. We then define the signal wave as being on the $xy$-plane, and describe its position with respect to the pump wave with an angle $\phi_s$. Finally, the position of the idler wave can be described using two angles: $\phi_i$, the angle of the idler wave with respect to the pump wave on the $xy$-plane, and $\theta_i$, the angle of the idler wave with respect to the pump wave on the $z$-axis. With this definition, the components $k_x$, $k_y$, and $k_z$ are given by
\begin{align}
    \Delta k_x &= 2k_p - k_s\cos{\phi_s} - k_i\cos{\phi_i}\cos{\theta_i} \\
    \Delta k_y &= k_s\sin{\theta_s} + k_i\sin{\phi_i}\cos{\theta_i} \\
    \Delta k_z &= k_i\sin{\phi_i},
\end{align}
where the signs are chosen in accordance with Fig.~\ref{fig:3D_spherical}.

\begin{figure}[htbp]
    \centering
    \begin{subfigure}{\textwidth}
    \centering
        \includegraphics[width = \textwidth]{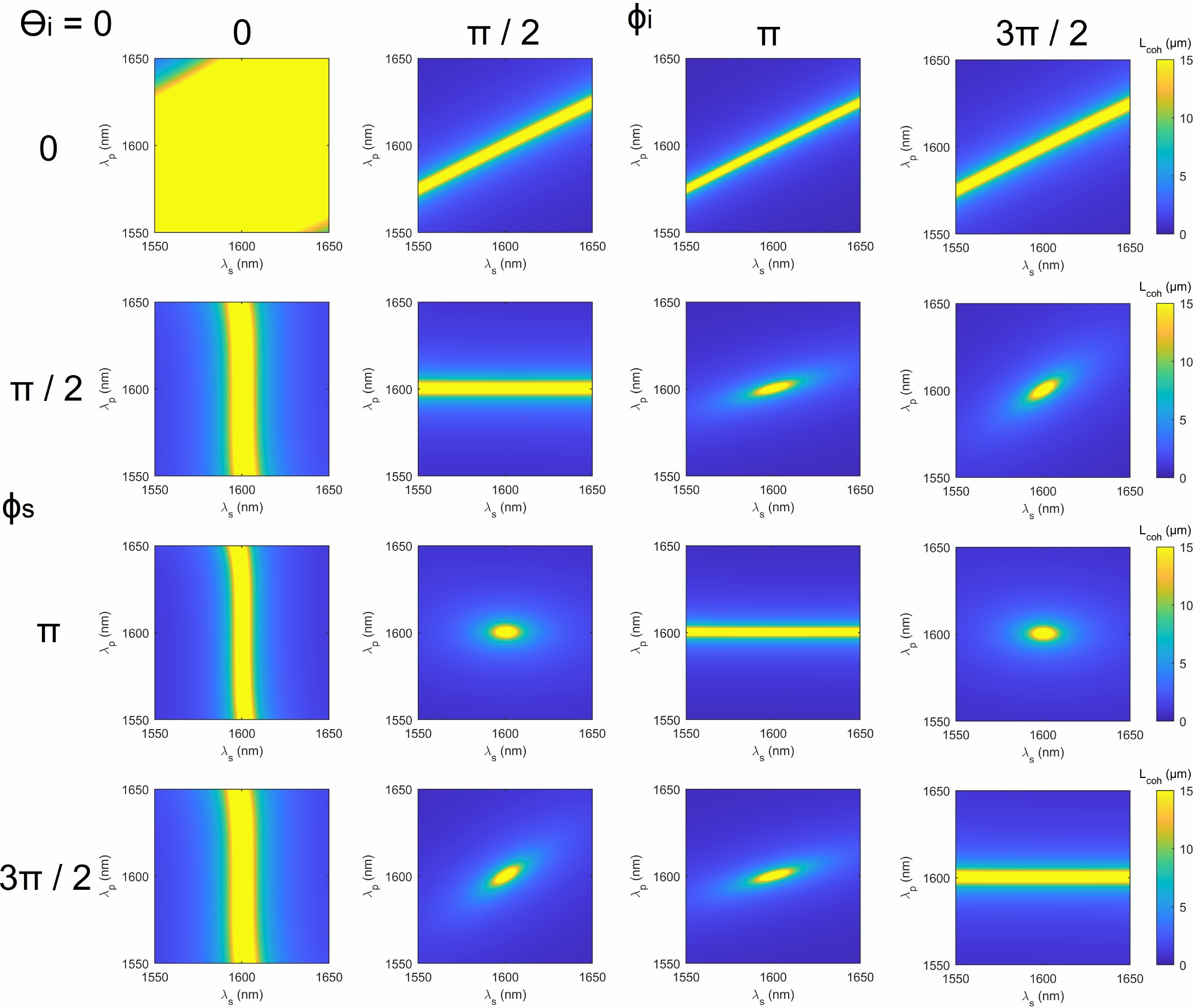}
    \end{subfigure}
    \begin{subfigure}{\textwidth}
    \centering
        \includegraphics[width = \textwidth]{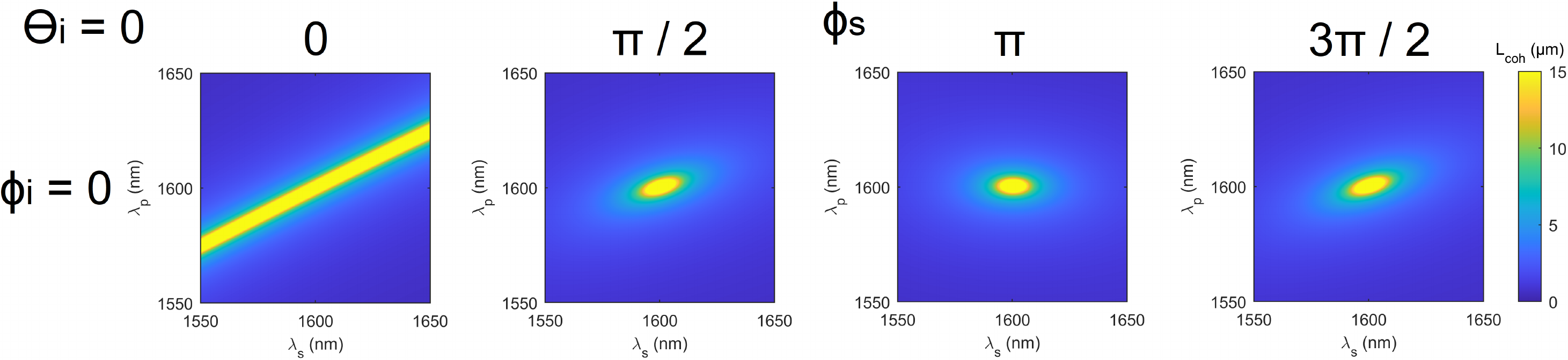}
    \end{subfigure}
    \caption{{\bf Numerical predictions of the coherence length.} The coherence lengths are calculated using $L_{\mathrm{coh}} = 2 / \Delta k$, for 20 separate combinations of the angles $\phi_s$, $\phi_i$, and $\theta_i$. As the angle of $\phi_i$ does not matter for $\theta_i = \frac{\pi}{2}$ (when $\theta_i = \frac{\pi}{2}$, the idler wave is entirely in the $z$-direction), we only include $\phi_i = 0$ for this case to avoid repetition. $L_{\mathrm{coh}}$ is plotted as a function of the pump wavelength $\lambda_p$ and the signal wavelength $\lambda_s$. $L_{\mathrm{coh}}$ is plotted from 0 \micron{} to 15~\micron, and the angles are given in radians.}
    \label{fig:results_arbitrary}
\end{figure}

After substituting these terms into Eq.~(\ref{delta_k_arbitrary}), we obtain an expression which simplifies to
\begin{equation}
	|\vec{\Delta k}| = \sqrt{4k_p^2 + k_s^2 + k_i^2 - 4k_p(k_s\cos{\phi_s} + k_i\cos{\phi_i}\cos{\theta_i}) + 2k_sk_i\cos{\theta_i}(\cos{\phi_s}\cos{\phi_i} + \sin{\phi_s}\sin{\phi_i})}.
	\label{eq:phase_match_arbitrary}
\end{equation}
From this expression, we can determine the phase-matching constraint for an arbitrarily oriented set of beams. This phase-matching condition, as a result, is a generalization of the phase-matching conditions in the main text. To get some intuition as to the results of this equation, we can plot a small subset of all the possible angles using the dispersion profile of our zero-index waveguide. In Fig.~\ref{fig:results_arbitrary}, we plot $L_\mathrm{coh}$ for the angles $\phi_s, \phi_i=0, \pi/2, \pi, 3\pi/2$ for the angles $\theta_i = 0$, $\pi/2$.

The subset of angles plotted in Fig.~\ref{fig:results_arbitrary} provides some intuition on the nature of Eq.~(\ref{eq:phase_match_arbitrary}) and on our claims regarding phase matching when $n = 0$. In the case where all angles are equal to zero, corresponding to co-propagation of all waves, we have phase matching everywhere. If only $\phi_s$ changes, we can achieve phase matching by simply placing the signal wave at the zero-index wavelength, and thereby eliminate the directional dependence of the displaced signal wave. This is the case for a forward-propagating idler wave with counter-propagating pump and signal beams, as seen in Fig.~\ref{fig:co-propagating_oop}a. We can analogously do the same when there is a variation in $\phi_i$ or $\theta_i$ for the idler wave (corresponding to a backward-propagating idler wave with co-propagating pump and signal beams, as seen in Fig.~\ref{fig:example_spectra_backward_forward}). If $\phi_s$ and $\phi_i$ are both altered equally, we can achieve phase matching by placing the \textit{pump} wave at the zero-index wavelength corresponding to a backward-propagating idler wave with counter-propagating pump and signal beams (as seen in Fig.~\ref{fig:co-propagating_oop}a).

There are cases where simply placing one of the components of the FWM interaction at the zero-index wavelength to enable phase matching does not work because all three constituent beams travel in seperate directions. This is the case for our demonstration with a signal beam incident from outside the plane of the device layer (Fig.~\ref{fig:co-propagating_oop}b). In such cases, we see that $L_{\mathrm{coh}}$ is large when all of the waves are clustered near the zero-index wavelength. While the phase-matching condition does vary with beam configuration, a FWM interaction for an arbitrary beam configuration is always phase-matched under some condition. This constitutes a theoretical prediction of phase matching which is free of directional restriction in low-index waveguides.

\section{Phase-matching nonlinear scattering theory}
\label{sec:simulations}

Using nonlinear scattering theory~\citeSM{obrien20152}, we qualitatively demonstrate both forward and backward-phase-matching in a zero-index medium consisting of a 2D Dirac cone photonic crystal. This method can be used to estimate the magnitude of the nonlinear signal generated in an interaction as a function of propagation length in, and thereby allowing for the direct estimate of the phase mismatch or coherence length. A benefit of this approach is that a realistic structured medium can be incorporated, and the effects of propagation loss or dispersion can be neglected. Therefore, these results can help us infer the contribution of phase matching.

\begin{figure}[H]
    \includegraphics[height=1.5in]{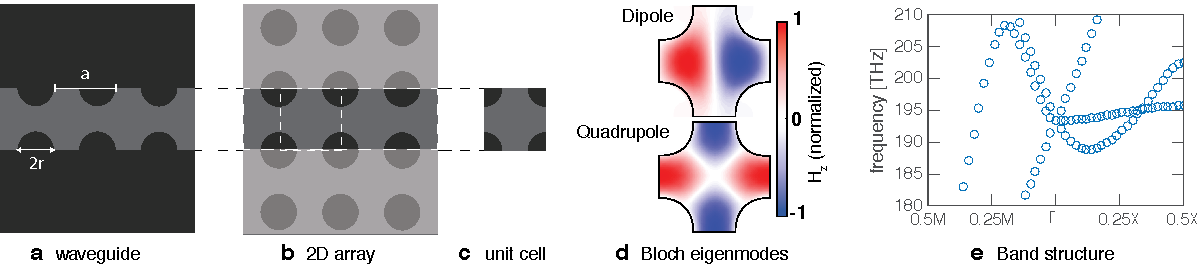}
    \centering
    \caption{{\bf Specifications of the low-index waveguide.} The design for the low-index waveguide {\bf a)} is derived from a 2D photonic crystal {\bf b)}. We model the unit cell from this array with periodic boundary conditions {\bf c)}, and obtain {\bf e)} a band structure that has a Dirac cone at the center of the Brillouin zone consisting of dipole and quadrupole modes {\bf d)}.}
    \label{fig:structure}
\end{figure}

The platform we investigated consists of a 2D square array of air holes in a silicon bulk ($a = 583$~nm, $2r = 364$~nm). The low-index waveguide probed in our measurements is a descendent of this theoretical metamaterial (Figs.~\ref{fig:structure}a-c) which is designed to exhibit a Dirac cone at the center of its Brillouin zone. The modes at the $\Gamma$-point consist of a dipole and a quadrupole mode (Figs.~\ref{fig:structure}d-e).

Nonlinear scattering theory predicts that the intensity of the nonlinear idler wave generated within a medium $I_\mathrm{NL}$ is proportional to the modulus square of overlap between an electric field originating from the detector at the idler frequency $\vec{E}$ and the nonlinear polarization induced by the source $\vec{P}^\mathrm{NL}$:
\begin{equation}
    I_{\textrm{NL}}\propto\left|\int\vec{E}_{\mathrm{detector}}\cdot\vec{P}_{\textrm{NL}}\,dV\right|^{2}.
\end{equation}

\begin{figure}[H]
    \includegraphics[height=1.5in]{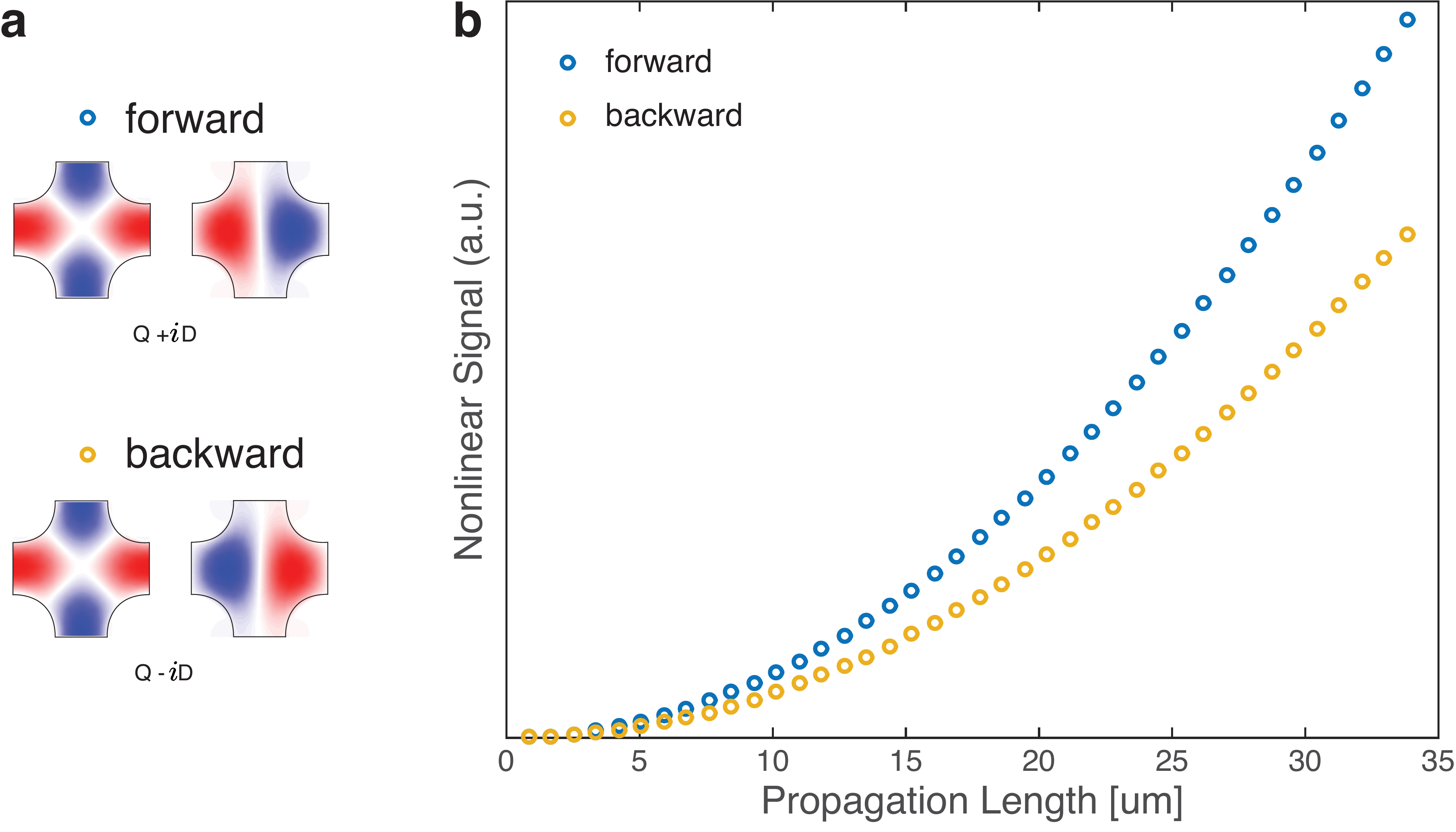}
    \centering
    \caption{{\bf Nonlinear signal generated in the zero-index medium as a function of propagation length.} For a perfectly phase-matched process, the generated signal for a forward-propagating idler wave is shown in blue, while the backward-propagating wave is shown in yellow.}
    \label{fig:nonlinear_scattering_theory}
\end{figure}

For a four-wave-mixing interaction, the nonlinear polarization $\vec{P}^{\mathrm{NL}}$ can be calculated using the mode distribution at the pump and signal frequencies $\vec{P}^{\mathrm{NL}} = \chi^{(3)} \vec{E}(\omega_p) \vec{E}^*(\omega_p) \vec{E}^*(\omega_s)$; Lorentz reciprocity dictates that the detector electric field $\vec{E}_{\mathrm{detector}}$ must be at the idler frequency. For a forward-propagating idler wave, the distribution of the detector field is $\vec{E}_{\mathrm{detector}} = \vec{E}_Q + i\vec{E}_D$, where $\vec{E}_Q$ and $\vec{E}_D$ represent the field distributions of the quadrupole and dipole modes, respectively. By contrast, a backward-propagating idler wave will have a field of $\vec{E}_{\mathrm{detector}} = \vec{E}_Q - i\vec{E}_D$ (Fig.~\ref{fig:nonlinear_scattering_theory}a).

Using the method outlined above, we calculate the nonlinear signal generated in the zero-index medium as a function of propagation length (Fig.~\ref{fig:nonlinear_scattering_theory}b). Our calculation simulates $I_\mathrm{NL}$ for a 2D array, which eliminates any out-of-plane radiative losses. Additionally, the calculation is performed at degenerate frequencies ($\omega_p = \omega_s = \omega_i = \omega$). The generated intensities are observed to grow quadratically in all propagation directions, indicating perfect phase-matching, consistent with a refractive index of zero. Additionally, the conversion efficiency for the backward-phase-matched signal is observed to be smaller due to a reduction in the total mode overlap caused by the out-of-phase dipole mode.

\section{Generation of the theoretical curves and loss profile}
\label{sec:loss_derivation}

While the coherence length provides us with some strong intuition as to when an interaction is phase-matched, it is alone insufficient to predict the generated power of a nonlinear interaction. To adequately analyze experimental results, we require a model that takes dispersion and radiative losses into account. We derive this model by solving the wave equation for each component of the field~\citeSM{boydnonlinearoptics2}. The wave equation is given by
\begin{equation}
	\nabla^2{\textbf{E}}_n - \frac{\epsilon_n\mu_n}{c^2}\frac{\partial^2{\textbf{E}}_n}{\partial t^2} = \frac{\mu_n}{\epsilon_0c^2}\frac{\partial^2{\textbf{P}}_n^{NL}}{\partial t^2}.
	\label{eq:wave_equation_final}
\end{equation}
where ${\textbf{E}}_n$ is the electric field of the incident light, ${\textbf{P}}_n^{NL}$ is the nonlinear component of the polarization density, $\epsilon_0$ is the vacuum permittivity, $\epsilon_n$ is the relative permittivity, $\mu_n$ is the relative permeability, and $c$ is the speed of light. Here, the subscript $n$ is used to denote an individual component of the field (\emph{i.e.,} pump, signal or idler). Given the scalar field approximation, we have the dispersion relation
\begin{gather}
	k_n = \frac{n_n\omega_n}{c}, \label{eq:k_omega_relation}\\
	n_n^2 = \epsilon_n\mu_n. \label{eq:n_epsilon_relation}
\end{gather}
where $k_n$ is the wave vector, $\omega_n$ is the angular frequency, and $n_n$ is the refractive index. For each component of the field, we can substitute a trial solution for the electric field and polarization densities
\begin{gather}
	{\textbf{E}}_n(z,t) = A_ne^{i(k_nz-\omega_nt)} + \textrm{c.c.} \\
	{\textbf{P}^\mathrm{NL}}_n(z,t) = P^\mathrm{NL}_ne^{-i\omega_nt} + \textrm{c.c.},
\end{gather}
where $A_n$ is the scalar amplitude of the electric field, and $P^\mathrm{NL}_n$ is the scalar amplitude of the nonlinear component of the polarization density. Here, c.c. represents the complex conjugate. The values $P^\mathrm{NL}_n$ for the pump, signal, and idler respectively (denoted by the subscripts $p$, $s$, and $i$) are given by
\begin{align}
	P^\mathrm{NL}_p &= 3\epsilon_0\chi^{(3)}A_p^*A_sA_ie^{i(k_sz+k_iz-k_pz)} + \textrm{c.c.}\\
	P^\mathrm{NL}_s &= 3\epsilon_0\chi^{(3)}{A_p}^2A_i^*e^{i(2k_pz-k_iz)} + \textrm{c.c.}\\
	P^\mathrm{NL}_i &= 3\epsilon_0\chi^{(3)}{A_p}^2A_s^*e^{i(2k_pz-k_sz)} + \textrm{c.c.}
\end{align}
Here, $\chi^{(3)}$ is the third-order nonlinear susceptibility. Solving the wave equation for every component of the FWM interaction yields the 3 coupled-amplitude equations
\begin{align}
    \frac{dA_p}{dz} &= \frac{3i\mu\chi^{(3)}\omega_p^2}{2k_pc^2}A_p^*A_sA_ie^{-i\Delta kz}, \\
	\frac{dA_s}{dz} &= \frac{3i\mu\chi^{(3)}\omega_s^2}{2k_sc^2}A_p^2A_i^*e^{i\Delta kz}, \\
	\frac{dA_i}{dz} &= \frac{3i\mu\chi^{(3)}\omega_i^2}{2k_pc^2}A_p^2A_s^*e^{i\Delta kz}.
\end{align}
To account for an interaction where the constituent waves are depleted by loss, we add a loss term proportional to the amplitude in the final result to represent the fields depleting as the waves propagate through the waveguide. For a forward-propagating idler wave, the coupled-amplitude equations, therefore, take the form:
\begin{align}
	\frac{dA_p}{dz} &= \frac{3i\mu_p\chi^{(3)}\omega_p}{2n_pc}A_p^*A_sA_ie^{-i\Delta k_\textrm{f}z} - \alpha_pA_p \label{eq:CA_pump_loss_co} \\
	\frac{dA_s}{dz} &= \frac{3i\mu_s\chi^{(3)}\omega_s}{2n_sc}A_p^2A_i^*e^{i\Delta k_\textrm{f}z} - \alpha_sA_s \label{eq:CA_signal_loss_co} \\
	\frac{dA_i}{dz} &= \frac{3i\mu_i\chi^{(3)}\omega_i}{2n_ic}A_p^2A_s^*e^{i\Delta k_\textrm{f}z} - \alpha_iA_i. \label{eq:CA_idler_loss_co}
\end{align}
Here, $\alpha_n$ is the propagation loss, and $\Delta k_\textrm{f} \equiv 2k_p - k_s - k_i$. Note that we have used the slowly varying amplitude approximation, and ignored the second derivative of $A_i$. One can obtain the parameter $\alpha$ from the propagation loss in dB/\micron by using~\citeSM{agrawalnonlinearfiberoptics2}
\begin{equation}
    \alpha_{\mathrm{1/m}} = \frac{\ln{10}}{20}\alpha_{\mathrm{dB/\mu m}}.
\end{equation}

\begin{figure}[H]
    \includegraphics[width=0.55\textwidth]{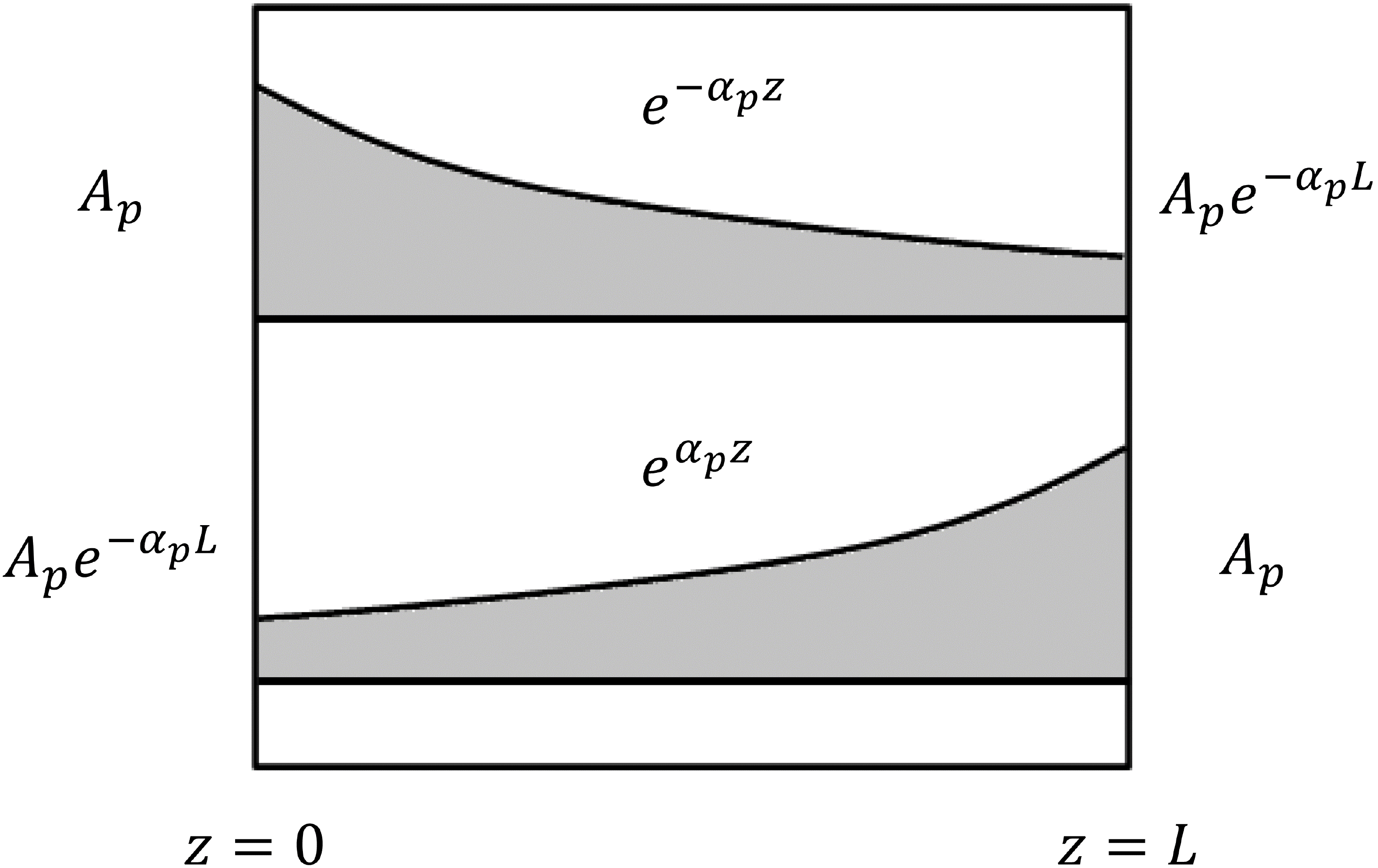}
    \centering
    \caption{{\bf Schematic demonstrating the swap in the frame of reference in the waveguide.} A pump wave propagating from $z = 0$ to $z = L$ is represented by a wave with initial amplitude $A_p$ that suffers loss $e^{-\alpha_pz}$ as it propagates through the waveguide, exiting with an amplitude $A_pe^{-\alpha_pL}$. In the treatment for the backward-propagating idler wave, we represent this same wave as propagating from $z = L$ to $z = 0$. We also appropriately invert the momentum $k_p$ of the pump wave.}
    \label{fig:backward_explanation}
\end{figure}

To solve these equations, we may use a relaxed version of the undepleted pump approximation. This relaxed approximation states that the depletion of the the pump and signal waves is dominated by propagation loss and not by conversion to idler waves. Using this approximation, we Eqs.~(\ref{eq:CA_pump_loss_co}~--~\ref{eq:CA_idler_loss_co}) to the form
\begin{gather}
    \frac{dA_p}{dz} = - \alpha_pA_p, \label{eq:CA_pump_loss_co_approx} \\
	\frac{dA_s}{dz} = - \alpha_sA_s, \label{eq:CA_signal_loss_co_approx} \\
	\frac{dA_i}{dz} = \frac{3i\mu_i\chi^{(3)}\omega_i}{2n_ic}A_p^2A_s^*e^{i\Delta k_\textrm{f}z} - \alpha_iA_i. \label{eq:CA_idler_loss_co_approx}
\end{gather}
These equations \textit{do} possess an analytical solution, as opposed to the previous ones. To solve them, we can first solve Eqs.~(\ref{eq:CA_pump_loss_co_approx}) and (\ref{eq:CA_signal_loss_co_approx}) to obtain an expression for the pump and signal wave amplitudes as a function of $z$. For the pump and signal waves, we obtain
\begin{gather}
    A_p(z) = A_{\textrm{p0}}e^{-\alpha_pz}. \\
    A_s(z) = A_{\textrm{s0}}e^{-\alpha_sz}.
\end{gather}
where $A_{\textrm{p0}}$ and $A_{\textrm{s0}}$ are the initial amplitudes of the pump and signal waves. After obtaining these expressions for the pump and signal waves, we can thereafter substitute them into Eq.~(\ref{eq:CA_idler_loss_co_approx}) to obtain
\begin{equation}
    \frac{dA_i}{dz} = \frac{3i\mu_i\chi^{(3)}\omega_i}{2n_ic}A_{\textrm{p0}}^2A_{\textrm{s0}}^*e^{i\Delta k_\textrm{f}z}e^{-\Delta\alpha z} - \alpha_iA_i \label{eq:diff_eq_idler_fw}
\end{equation}
where we have defined $\Delta\alpha \equiv 2a_p + a_s$ for notational convenience. We are now left with a single linear differential equation which can be solved. By using the initial condition $A_i(0) = 0$, we obtain the final expression for the idler wave amplitude as a function of $z$
\begin{equation}
    A_i(z) = \frac{3i\mu_i\chi^{(3)}\omega_i}{2n_ic}A_{\textrm{p0}}^2A_{\textrm{s0}}^* \left(\frac{e^{(i\Delta k_\textrm{f} - \Delta\alpha)z} - e^{-\alpha_iz}}{i\Delta k_\textrm{f} - \Delta\alpha + \alpha_i}\right). 
    \label{eq:diff_eq_idler_fw_final}
\end{equation}
This equation can be used to calculate the power of a forward-propagating idler wave as a function of the propagated length $z$ in the waveguide. We can follow an analogous procedure to obtain the expression for a backward-propagating idler wave. In the backward-propagating case, the phase-matching condition is $\Delta k_\textrm{b} = 2k_p - k_s + k_i$. As the idler wave is counter-propagating against the pump and signal, we can define our pump and signal waves to begin at $z = L$. This effectively inverts the frame of reference of the waveguide (See Fig.~\ref{fig:backward_explanation} for an illustration clarifying this). We also appropriately invert the signs of the pump, signal, and idler wave momentum terms. In cases where the relaxed undepleted pump approximation holds, this approach is theoretically valid. Under these assumptions, the coupled-amplitude equations take the form

\begin{gather}
    \frac{dA_p}{dz} = \alpha_pA_p, \label{eq:CA_pump_loss_cobw_approx} \\
	\frac{dA_s}{dz} = \alpha_sA_s, \label{eq:CA_signal_loss_cobw_approx} \\
	\frac{dA_i}{dz} = \frac{3i\mu_i\chi^{(3)}\omega_i}{2n_ic}A_p^2A_s^*e^{-i\Delta k_\textrm{b}z} - \alpha_iA_i. \label{eq:CA_idler_loss_cobw_approx}
\end{gather}
Using the initial conditions $A_p(0) = A_{\mathrm{p0}}e^{-\alpha_pL}$ and $A_s(0) = A_{\mathrm{s0}}e^{-\alpha_sL}$, we can obtain the pump and signal amplitudes as a function of $z$, obtaining
\begin{align}
    A_p(z) &= A_{\textrm{p0}}e^{-\alpha_pL}e^{\alpha_pz}, \\
    A_s(z) &= A_{\textrm{s0}}e^{-\alpha_sL}e^{\alpha_sz}.
\end{align}
By substituting these expressions for the pump and signal wave amplitudes into Eq.~(\ref{eq:CA_idler_loss_cobw_approx}) and subsequently solving the resulting linear differential equation, we obtain the result
\begin{equation}
    A_i(z) = \frac{3i\mu_i\chi^{(3)}\omega_i}{2n_ic}A_{\textrm{p0}}^2A_{\textrm{s0}}^* e^{-(2\alpha_p + \alpha_s)L}\left(\frac{e^{(\Delta\alpha - i\Delta k_\textrm{b})z} - e^{-\alpha_iz}}{\Delta\alpha - i\Delta k_\textrm{b} + \alpha_i}\right). \label{eq:diff_eq_idler_bw_final}
\end{equation}
We use Eqs.~(\ref{eq:diff_eq_idler_fw_final}) and (\ref{eq:diff_eq_idler_bw_final}) to analyze our experimental results. The power $P$ of a wave (not to be confused with the polarization density $P^{\mathrm{NL}}$), which we measure is proportional to the square of the modulus of the field amplitude: $P \propto n{|A|}^2$. In the power equation, we use the refractive index of the surrounding silicon waveguide which couples into the zero-index waveguide where the measurement is performed.

\begin{figure}[H]
    \includegraphics[width=0.6\textwidth]{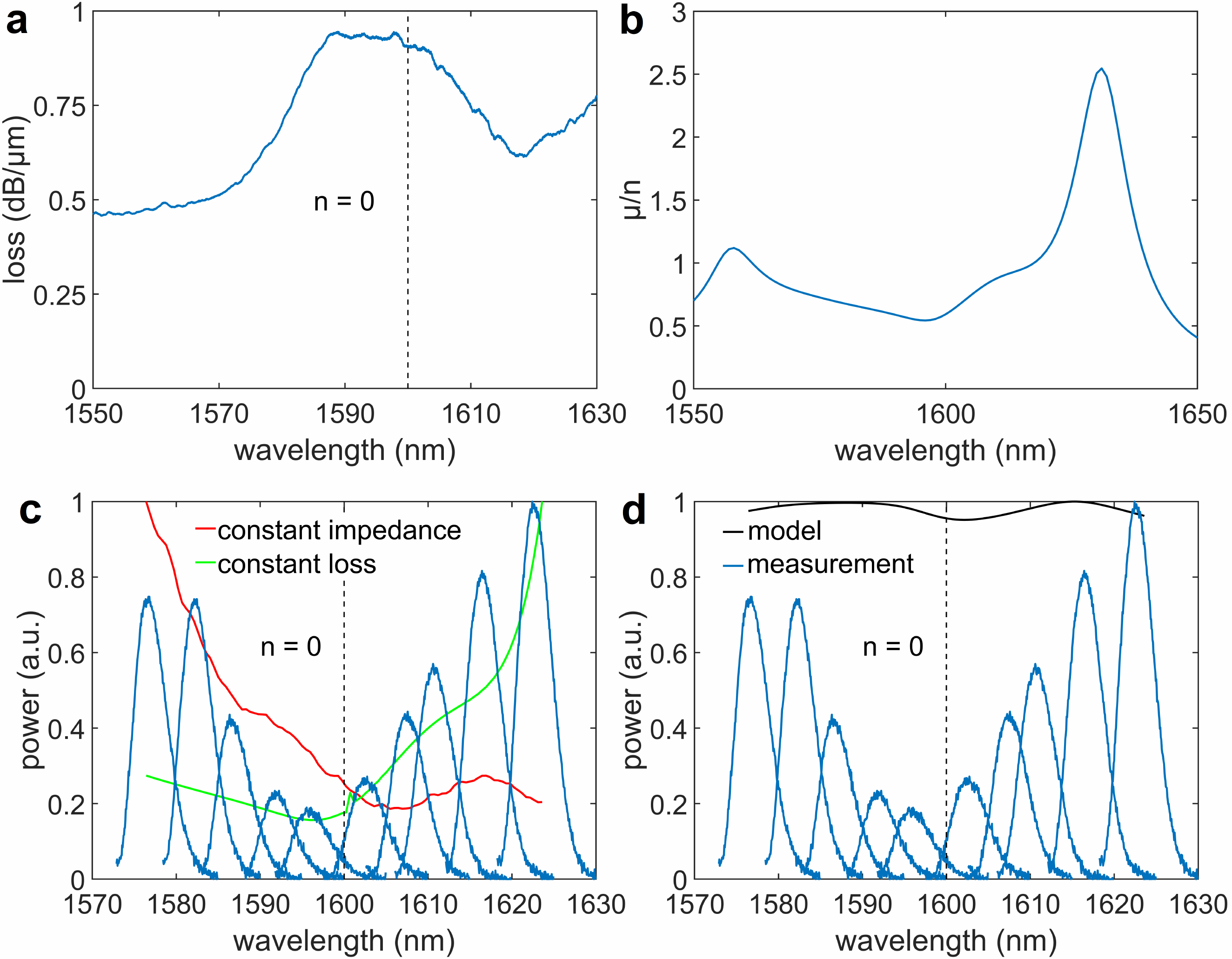}
    \centering
    \caption{{\bf Loss and permeability over $n$ values and their effects of the produced idler peaks.} {\bf a)}~Loss profile of a similar waveguide in dB/μm. The loss peaks shortly below the zero-index wavelength (dashed black line). {\bf b)}~Simulated $\mu/n$, a quantity proportional to the impedance, plotted as a function of wavelength. This quantity is included as a factor in Eqs.~\ref{eq:diff_eq_idler_fw_final} and \ref{eq:diff_eq_idler_bw_final}. {\bf c)}~Generated idler wave spectra in the forward-propagating direction for a FWM process with co-propagating pump and signal beams, compared to two theoretical predictions. All results are normalized to unity. For the red curve, the impedance is made wavelength-invariable, allowing us to isolate the effects of loss. For the green curve, the loss is made wavelength-invariable, allowing us to isolate the effects of impedance. While the red curve predicts an attenuation beyond the dip in the forward-propagating idler peaks, the green curve predicts an attenuation prior to the dip in the forward-propagating idler peaks. {\bf d)}~The same experimental data plotted against the theoretical prediction (black curve) when all wavelength-dependant quantities bar the phase matching condition are held constant.}
    \label{fig:impedance_loss}
\end{figure}

When based solely on phase-matching constraints, our discussion in Section~\ref{sec:unrestricted} suggests that the forward-propagating idler peaks should have equal magnitudes at every wavelength (Fig.~\ref{fig:impedance_loss}d). However, our measurements results indicate that this is clearly not the case. This discrepancy can be explained by incorporating a dispersive loss and permeability to the model (Fig.~\ref{fig:impedance_loss}a--b). These factors are included in the model that generated the theoretical curves plotted in Fig.~\ref{fig:example_spectra_backward_forward} of the main text.

In Fig.~\ref{fig:impedance_loss}c, we isolate the effects of these two quantities on the output idler power by plotting superimposing two curves over the measurement results: a constant permeability and therefore a constant impedance with a variable loss (red curve); and a variable permeability and a constant loss (green curve). The propagation loss values were extracted from previous measurements on a similar set of waveguides, and the permeability was extracted from simulations. When the impedance is held constant, the forward-propagating idler is attenuated for wavelengths longer than 1600~nm. When the loss is held constant, the forward-propagating idler is attenuated for wavelengths shorter than 1600~nm. When both curves are compared, it is clear that the spectrum of the forward-propagating idler is a result of the combination of these two factors. As a result, we can conclude that the spectrum of the forward-propagating idler peaks is caused by wavelength-variable impedance and loss, while the spectrum of the backward-propagating idler peaks is primarily the result of the phase-matching condition.

In summary, the fact that the simultaneously generated forward and backward-propagating idler peaks respond so differently to passing through the zero-index wavelength is overall strong proof that the idler wave both show directional independence, and that each direction is subject to its own phase-matching requirement.

\section{Lower bound estimate on the coherence length}
\label{sec:lower_bound}

For collinear beams, the amplitude of the generated signal wave may be shown to be a function of the coherence length $L_{\mathrm{coh}} = 2/\Delta k$ and interaction length $L$~\cite{boydnonlinearoptics}: 
\begin{equation}
    \label{eq:amplitude}
	A_i \propto L\sinc{(L/L_{\mathrm{coh}})},
\end{equation}
where we assume the slowly varying amplitude approximation, and neglect the propagation losses. Here, we define the lower bound on the coherence length as the shortest length for which the factor $\sinc^2{(L/L_{\mathrm{coh}})}$ (obtained by squaring Eq.~\ref{eq:amplitude}) yields a pulse at a quarter of the power of the peak power. As the power is proportional to the square of the amplitude, we may write
\begin{equation}
	P_i \propto \sinc^2{(L/L_{\mathrm{coh}})}.
\end{equation}
The solution to $\sinc^2{(x)} = 0.25$ is $x = 1.895$. Therefore, for a 14.8~\micron{} waveguide, the lower bound of the coherence length is given by $L_{\mathrm{coh}} = 14.8~$\micron{}$ / 1.895 = 7.81$~\micron{}. If we, instead, use a lower power to define our lower bound, the lower bound on the coherence length is longer. $7.81~\micron{}$ is dramatically longer than a free-space wavelength, and proves we have phase-matching beyond what is possible in a metasurface.

\section{Waveguide dispersion}
\label{sec:dispersion}

\begin{figure}[H]
    \centering
    \includegraphics[width=0.5\linewidth]{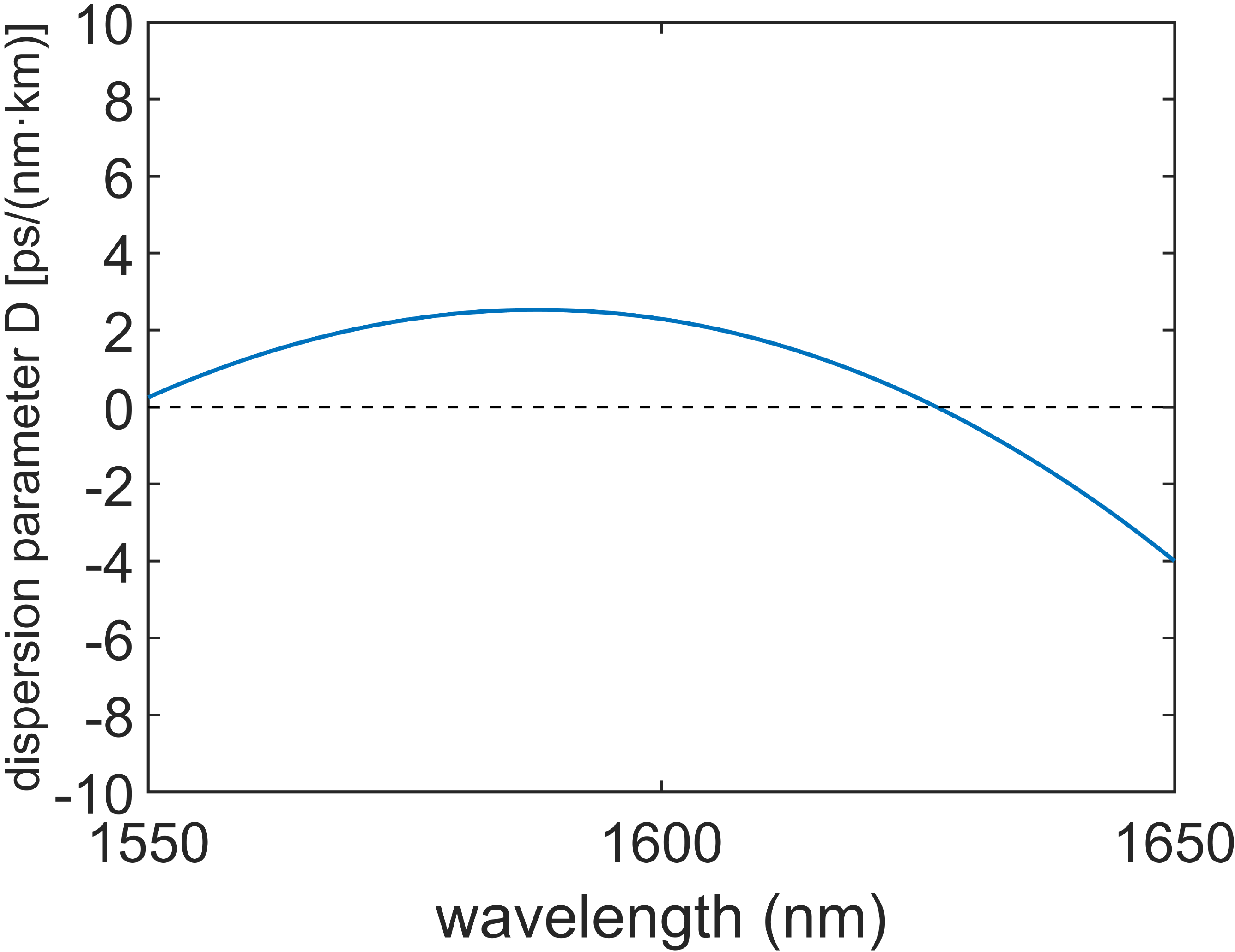}
    \caption{{\bf Simulated dispersion parameter D for a Dirac-cone zero-index waveguide.} D is positive, corresponding to anomalous dispersion over the 80~nm surrounding the zero-index wavelength.}
    \label{fig:dispersion}
\end{figure}

The waveguides used in the experiment possess anomalous dispersion in a bandwidth of roughly 80~nm surrounding the zero-index wavelength (Fig.~(\ref{fig:dispersion})). This anomalous dispersion ensures that the desired FWM nonlinear optical process is phase-matched in the spectral region of interest, as elaborated upon in \citeSM{foster20062}.

\section{Experimental setup}
\label{sec:setup}

\begin{figure}[H]
    \includegraphics[width=\textwidth]{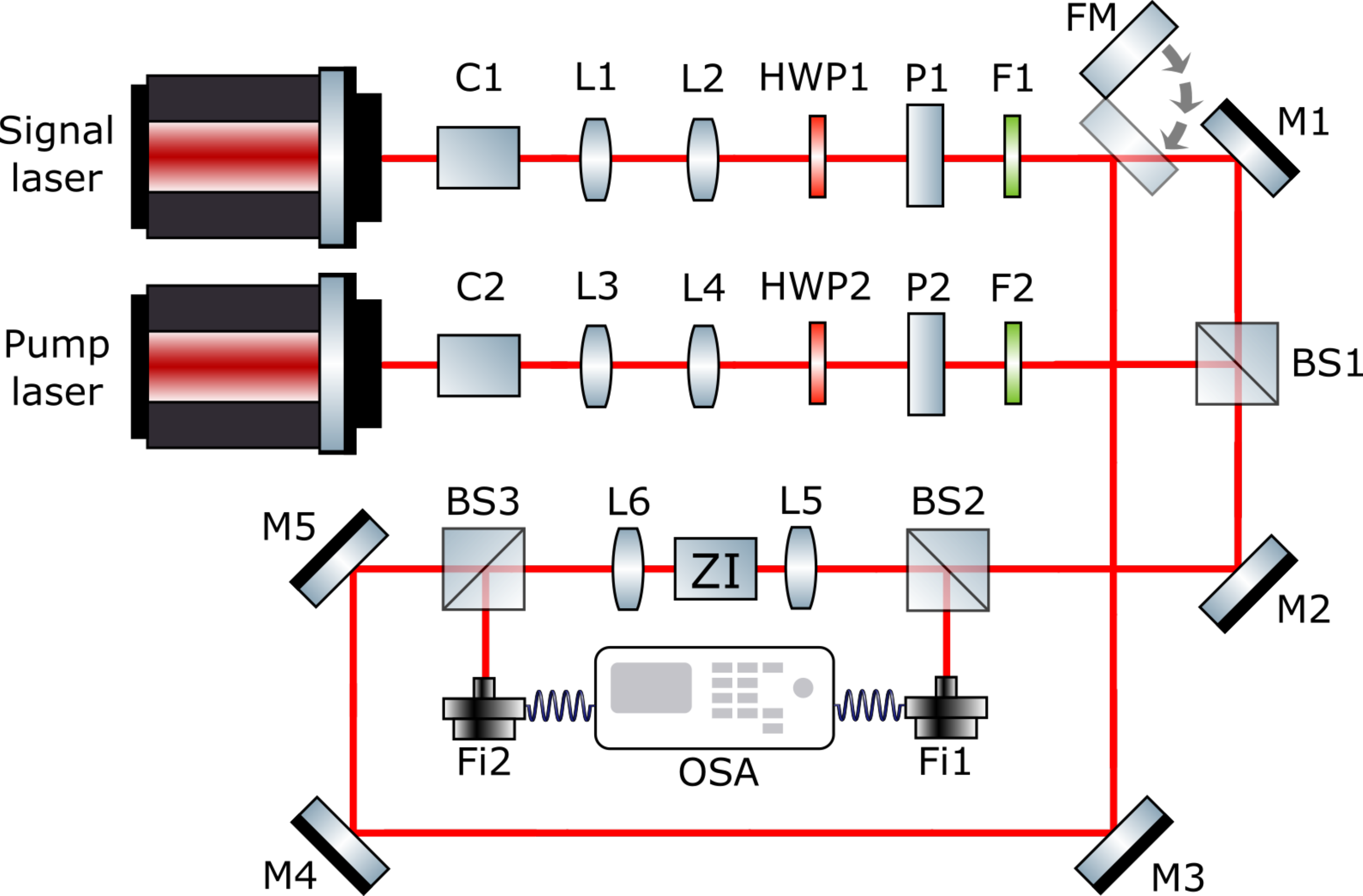}
    \centering
    \caption{{\bf Schematic of the setup used to couple the pump and signal beams into the zero-index waveguide.} The pump and amplified signal beams pass through lens systems to collimate the beams (C1 and C2), and have their spot sizes matched using telescopes (L1, L2, L3, L4). The beam is subsequently polarized using a half-wave plate and polarizing beam-splitter (HWP1, HWP2, P1, P2), and then spectrally filtered (F1, F2). In the case of co-propagating pump and signal beams, the beams are combined using a beamsplitter cube (BS1) and a lens (L5) subsequently focuses the beams into the zero-index waveguide (ZI). Optical fibers (Fi1, Fi2) on either side collect the forward and backward-propagating generated light exiting the waveguide, and send it to the OSA for further analysis. For counter-propagating beams, the flip mirror (FM) is flipped into position, and the signal light is focused into the zero-index waveguide via the lens L6. Mirrors which redirect light are represented by M.}
    \label{fig:optical_setup}
\end{figure}

A schematic of the complete setup can be seen in Fig.~\ref{fig:optical_setup}. Both beams must be collimated with roughly the same spot size (0.5~cm in diameter) so that co-propagating beams can be coupled into the zero-index waveguide via the same lens. We use a system of lenses to collimate the beams, and we use a telescope to adjust the spot size while retaining the collimation. To polarize the lasers, we use the combination of a half-wave plate and polarizing beam-splitter for each laser to both achieve the desired transverse-electric polarization and provide a means with which we can modulate the power of the pump and signal beams. 

Following polarization, the pump and signal beams are coupled into the zero-index waveguide. In the case of co-propagating beams, we combine the pump and signal beams using a beamsplitter cube (beam cube). The portion of the pump and signal beams that is not used to couple into the waveguide is then directed towards a detector to determine the power of the pump and signal beams when performing measurements. Once combined, the pump and signal beams can be coupled into the waveguide through its coupling pads.

To determine the generated output, we collect the light using multi-mode optical fibers on both sides on the waveguide. On the side of the input facet, we set up a non-polarizing beam cube to allow the input light to travel through while redirecting the output light to our detector. On the opposite side of the waveguide, we focus the output light on our multi-mode fiber. Once the light has been collected, it is spectrally decomposed by an AQ-6315E optical spectrum analyzer (OSA) for subsequent analysis.

\section{Idler wave power measurement}
\label{sec:power_measurement}

\begin{figure}[H]
    \centering
    \includegraphics[width=0.6\linewidth]{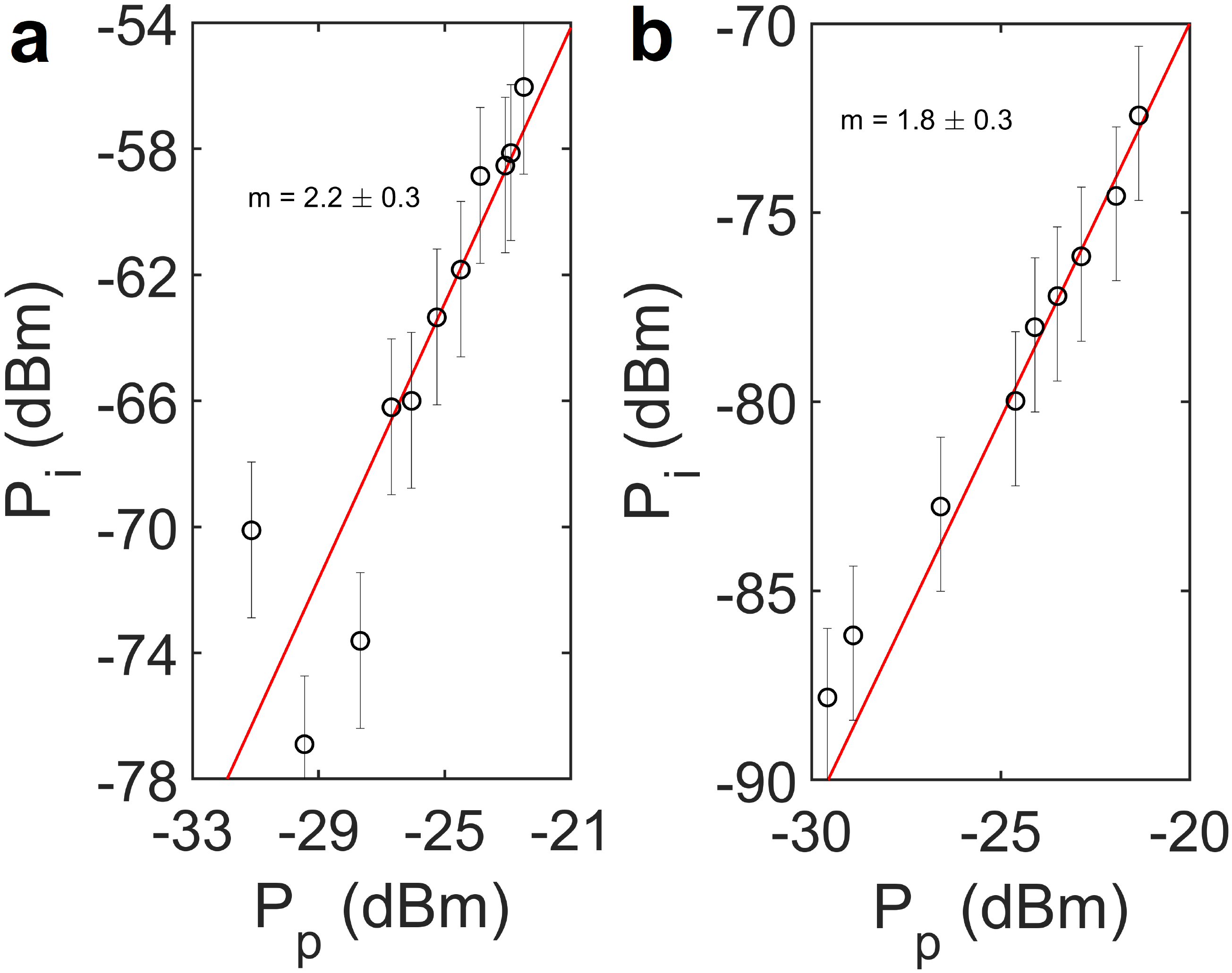}
    \caption{{\bf Peak power of the idler wave $P_i$ as a function of the peak power of the pump wave $P_p$.} In {\bf a)}, we plot the powers for forward-propagating light, while in {\bf b)}, we plot the powers for backward-propagating light. The error bars represent measurement uncertainties extracted from the standard deviation of a repeated set of measurements.}
    \label{fig:power_measurement}
\end{figure}

To confirm that the generated idler wave is the product of a third-order FWM interaction, we plot the peak power of the idler wave as a function of the peak power of the pump beam for both the forward and backward-propagating light (Fig.~\ref{fig:power_measurement}).

For the forward-propagating light, we observe a slope of $2.2 \pm 0.3$, while for the backward-propagating light, we observe a slope of $1.7 \pm 0.3$. Both of these values are close to 2, confirming the prediction of a quadratic relationship between pump and idler power. Deviations from 2 likely occur due to the background noise of the OSA and fluctuations in the pump spectrum. It can summarily be concluded that the idler wave is produced by a FWM process~\citeSM{boydnonlinearoptics2}.

\end{document}